\documentclass{elsart}
\def\cs{$^{137}$Cs}

\def\rn{$^{222}$Rn}
\def\Poa{$^{218}$Po}

\def\Poc{$^{210}$Po}

\def\Pbb{$^{210}$Pb}
\def\Pbc{$^{206}$Pb}
\def\Pbb{$^{210}$Pb}
\def\Bi{$^{210}$Bi}

\def\csitl{CsI(T$\ell$)~}

\def\Journal#1#2#3#4{{#1} {\bf #2}, #3 (#4)}

\def\NIM{Nucl. Instrum. Methods}

\def\PLB{Phys. Lett.  B}
\def\ASP{Astropart. Phys.}
\def\PRL{Phys. Rev. Lett.}

\def\EPJA{Eur. Phys. J. A}
\def\etal{{\it et al.}}

\usepackage{epsfig}

\usepackage{amssymb}
\usepackage{footnote}
\usepackage[bookmarks]{hyperref}

\begin{document}

\begin{frontmatter}

\title{Low energy fast events from radon progenies at the surface of a \csitl scintillator}

\author[snu]{S.C.~Kim,}\footnote{sckim@hep1.snu.ac.kr}
\author[snu]{H.~Bhang,}
\author[snu]{J.H.~Choi,}
\author[sejong]{W.G.~Kang,}
\author[knu]{H.J.~Kim,}
\author[snu]{K.W.~Kim,}
\author[snu]{S.K.~Kim,}\footnote{skkim@hep1.snu.ac.kr}
\author[sejong]{Y.D.~Kim,}
\author[snu]{H.S.~Lee,}
\author[snu]{J.I.~Lee,}
\author[snu]{J.H.~Lee,}
\author[snu]{J.K.~Lee,}
\author[snu]{M.J.~Lee,}
\author[snu]{S.J.~Lee,}
\author[snu]{J.~Li,}
\author[tsinghua]{J.~Li,}
\author[tsinghua]{Y.J.~Li,}
\author[snu]{X.~Li,}
\author[snu]{S.S.~Myung,}
\author[snu]{S.L.~Olsen,}
\author[snu]{S.~Ryu,}
\author[snu]{I.S.~Seong,}
\author[knu]{J.H.~So,}
\author[tsinghua]{Q.~Yue}
\center (\author{KIMS Collaboration})

\address[snu]{Department of Physics and Astronomy, Seoul National University, Seoul 151-742, Korea}
\address[sejong]{Department of Physics, Sejong University, Seoul 143-747, Korea}
\address[knu]{Physics Department, Kyungpook National University, Daegu 702-701, Korea}
\address[tsinghua]{Department of Engineering Physics, Tsinghua University,
Beijing 100084, China}

\begin{abstract}
In searches for rare phenomena such as elastic scattering of dark matter particles
or neutrinoless double beta decay,
alpha decays of \rn~ progenies attached to the surfaces of the detection material 
have been identified as a serious source of background.
In measurements with \csitl scintillator crystals,  we demonstrate that alpha decays 
of surface contaminants produce fast signals with a characteristic mean-time 
distribution that is distinct from those of neutron- and gamma-induced events.
\end{abstract}

\begin{keyword}
alpha \sep \rn \sep \csitl crystal \sep dark matter \sep pulse shape discrimination 
\end{keyword}
\end{frontmatter}

\section{Introduction}

\begin{figure}[!htp]
\center \psfig{figure=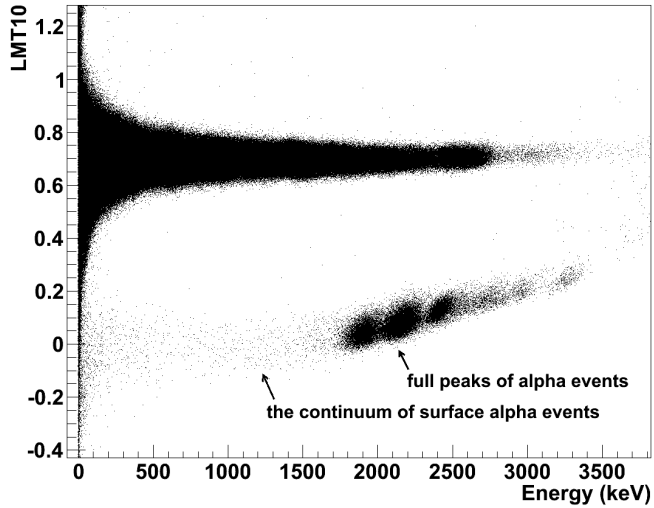, angle=0,width = 4.0 in}
\caption{Deposited energy (horizontal) versus LMT10 (vertical) in data from the KIMS experiment. 
 Here LMT10 denotes the natural logarithm of the mean-time of each event
 computed over first 10 $\mu$s interval.  Here one can see that SA events are present as a band whose LMT10
 is small compared to the background, which indicates that they decay quickly. } \label{sakims}
\end{figure}

 Signals caused by radioactive decays of \rn~ progenies that adhere to detector surfaces are 
 known to be serious background for rare phenomena 
 experiments such as searches for WIMPs (Weakly Interacting Massive Particles) dark matter 
 and neutrinoless double beta decay~\cite{anomalFE,cuoreA,cresstA}. 
 \rn~ is a noble gas that permeates the air. 
 It decays to \Poa, which is a reactive metal that readily adheres 
 to almost any surface~\cite{Rnmechanism}. Through successive decay 
 chains, the progenies of \rn~ can accummulate on detector surfaces. 
 When they decay, only a portion of the energy of the decay products is detected,
 thereby producing a troublesome low energy background. 
 Among the progenies of \rn, \Pbb~ and \Poc~ are the most dangerous because they have long half-lives, 
 22.3 years and 138 days, respectively. 
 If an experimental system is isolated,  \Poc~ becomes the principal alpha emitting contaminant. 
 The energy spectrum of these surface events appears as a continuum that ranges from the full peak energy of
 the alpha down to the very low energies.  
 Anomalously fast events have been observed in WIMP search experiments that use inorganic
 scintillators, and these have been attributed to 
 the effects of surface alpha (SA) events since their rate is seen to be reduced when the detector surfaces are polished~\cite{anomalFE,sa_study1,sa_study2}.
 The KIMS experiment at the Yangyang Underground Laboratory in Korea looks for WIMP-induced nuclear recoils in an array
 of \csitl scintillators~\cite{prllee}. 
 Figure~\ref{sakims} shows a scatter plot of energy deposits versus LMT10 for events seen in the KIMS detector,
 where LMT10 denotes the natural logarithm of the  mean-time of each event
 calculated for a 10 ${\mu}$s interval starting from the beginning of the event
 (MT10$= \frac{\sum_{t_i<10\mu s}{A_i\times t_i}}{\sum_{t_i<10\mu s}{A_i}}$, $A_i$ is the area of the $i^{\rm th}$ cluster,
 which is usually equivalent to a single photo-electron, of an event).
 In the lower right-hand portion of the plot, strong alpha-decay peaks can be seen with tails that extend down 
 to zero energy and low LMT10 values. 
 Here one can see that LMT10 values for the low energy tail events are distinctly
 smaller than those from gamma-ray induced signals.

WIMPs are expected to scatter elastically from nuclei. The KIMS experiment is designed to detect the energy
deposited by the recoiling nucleus and uses a pulse shape discrimination (PSD) analysis to distinguish nuclear
recoil events from  gamma-induced
backgrounds~\cite{prllee}.    The presence of surface alpha events at energies below 10 keV, the main region of interest for 
the dark matter search, complicates the PSD analysis.

\section{Experimental setup}

  In order to investigate the characteristics of SA background events, we contaminated the surface
  of a small \csitl sample crystal with \rn~ progenies
  by placing it for four days in a special chamber at the Korea Research Institute of Standards and Science (KRISS)
  in which the \rn~ concentration was around 4.33 MBq/m$^{3}$. 
  The size of the contaminated crystal is 3~cm $\times$ 3~cm $\times$ 1.4~cm.

\begin{figure}[!htp]
\center \psfig{figure=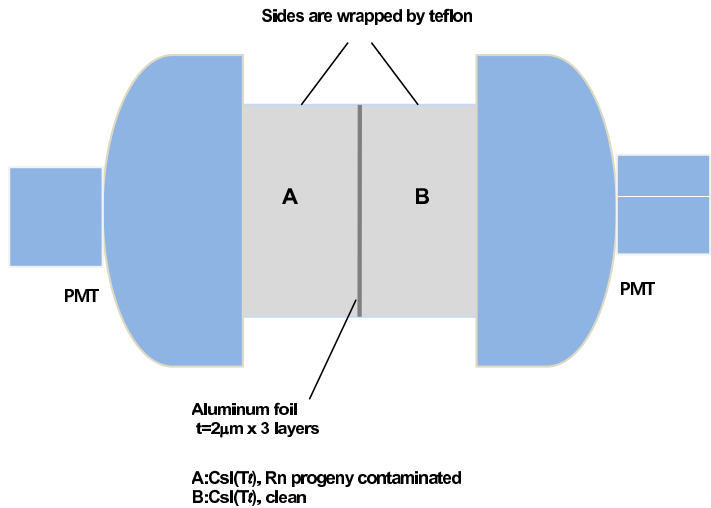,angle=-90, width = 4.0 in}
\caption{ A schematic depiction of the Radon progenies contaminated Double crystal Detector (RDD). Crystal A is the \rn -progenies contaminated \csitl crystal
and crystal B is a clean \csitl crystal that is used to detect an alpha particle that escapes from crystal A. 
The two crystals are separated by three 2 $\mu$m-thick layers of aluminum foil.  
} \label{RDD}
\end{figure}

  The parent \rn~ nucleus decays to \Poa~ with a half--life of 3.8 days. 
  In a period of several tens of minutes, the \Poa~ nucleus decays to \Pbb~ via several decay steps. 
  \Pbb~ beta decays with a half--life of 22.3 years to \Bi , which subsequently beta decays to \Poc. 
  From this series of beta decays, \Poc, the main source of surface alpha events, is continuously produced on the crystal surface.  
  It decays into \Pbc~ with a half-life of 138 days, emitting a 5304 keV alpha
  particle. The kinetic energy of the recoiling \Pbc~ nucleus is 103 keV.

\begin{figure}[!htp]
\center \psfig{figure=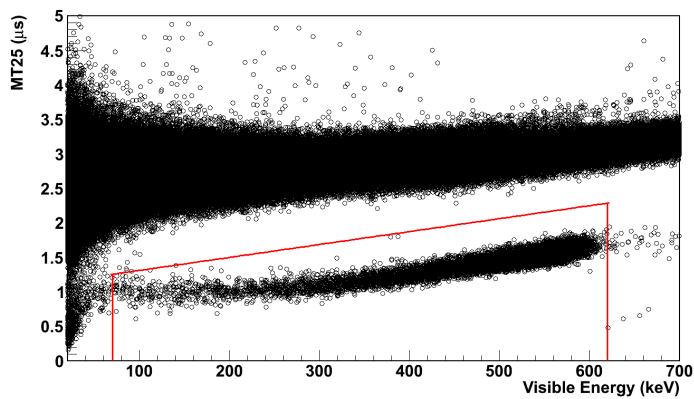,angle = 0, width = 4.0 in}
\caption{ MT25 versus the visible energy in crystal B, where MT25 is the mean-time of each event that 
 is determined over a 25 $\mu$s event-time window.
 The events inside the red solid lines are selected as alphas escaping from crystal A. 
}\label{alphapartB}
\end{figure}

  Figure~\ref{RDD} shows a schematic diagram of the experimental
  setup, which we named the Radon progenies contaminated Double crystal Detector (RDD), that was used to study the
  characteristics of SA events.  In order 
  to tag escaping surface alpha particles,  we attached a clean \csitl crystal (crystal B in Fig. \ref{RDD}) 
  to a face of the contaminated one (crystal A in Fig. \ref{RDD}). 
  We inserted three 2 $\mu$m-thick aluminum foils between the two crystals to provide a barrier
  to prevent cross-talk of the scintillation light between the crystals that the 5 MeV alphas can easily penetrate. 
  According to the SRIM program~\cite{SRIM}, 
  the energy loss of a 5304 keV alpha in a 6 $\mu$m-thick aluminum layer is 998 keV; thus alphas that penetrate the foils still have
  ample energy to provide a trigger signal.  
  These layers also serve as a reflector for the collection of scintillation light.
  The sides of the crystals are wrapped with teflon tape.
  A 3 inch photomultiplier tube (PMT), 9269QA from Electron Tubes, Ltd. is attached to
  each crystal to detect the scintillation photons. 
  The PMT signals are amplified 100 times by a fast amplifier from Notice Co., Ltd.
  The amplified signals are digitized by a 400 MHz Flash Analog-to-Digital Converter(FADC) of Notice Co., Ltd. that is mounted in a Versa Module
  Eurocard (VME) crate that is read-out by a linux-operating PC via a VME--USB2 interface. 
  The DAQ system is based on the ROOT package~\cite{root}.  
  When two or more photoelectrons (PEs) are detected in each PMT within a 2 $\mu$s time window, a trigger is generated. 
  Events with pulse width longer than 300 ns are also triggered in order to include high energy events in which many PEs are merged into a single big pulse. 
  For each event, the PMT responses throughout a 40.96 $\mu$s time window is recorded.
  Of these, pulses inside a 25 $\mu$s window are used for analysis. 

\begin{figure}[!htp]
\center \psfig{figure=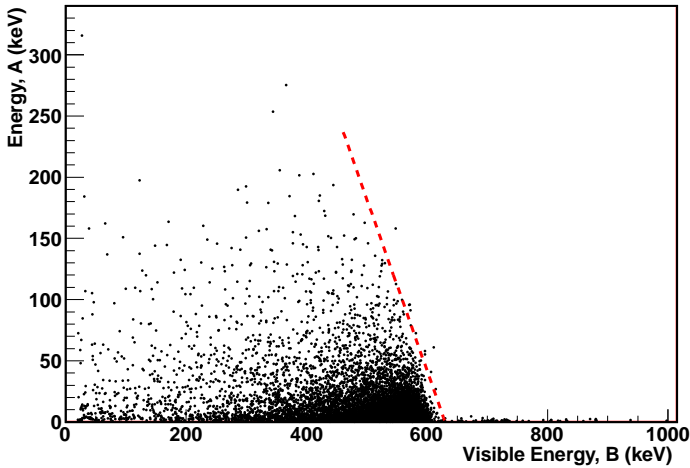,angle=0, width = 4.0 in}
 \caption{ Energy of crystal A versus the visible energy of crystal B for events that are tagged as an alpha.
}\label{EalphapartAnB}
\end{figure}

  To tag SA events in crystal A, we require a signal in crystal B that is consistent with that of an alpha particle. 
  Because of the good PSD power of \csitl scintillators, as shown in Fig.~\ref{alphapartB}, alpha signals 
  are clearly separated from other backgrounds. Here, MT25, the mean-time of each event
  determined over a 25 $\mu$s time window, is used. In Fig.~\ref{alphapartB}, 
  the visible energy is the measured energy which differs from the actual energy above around 100 keV because of the saturation effect
  resulting from the DAQ system optimized for collecting low energy events. Furthermore, the energy of alpha events 
  is underestimated more by the quenching factor in \csitl, which is about 0.5.
  No corrections for these effects have been applied.
  Events that populate the region inside the red solid lines of Fig.~\ref{alphapartB} are selected as alphas that escape from crystal A.
  Figure \ref{EalphapartAnB} shows a scatter plot of energy in crystal A versus the visible energy in crystal B for events tagged as
  alphas.
  This figure shows the sum of the energies at both crystals has an upper bound, as expected.
  The wide spread in the sum of the energies reflects the energy loss in the Al foil layers and possible dead layer on the surface of the detector
  ,which depends upon the alpha particle's direction.
  To separate SA events from background, an  upper bound of alpha energy at crystal B is set at 620 keV and a lower bound at 70 keV.  
  The experiment started two months after the radon contamination of the crystal was done and ran
  for about 3 months. The rate of events tagged as SA increased with time because of the continuous supply of \Poc~ from \Pbb.
  The event rate became 947 events$/$day by the end of the measurement period. 
  From this value, the amount of \Pbb~ contaminants implanted at the surface of the crystal is estimated to 
  be about $ 1.6 \times 10^8 $ and the adhesion rate of radon progenies in the radon chamber is estimated to be about 13$/$cm$^2/$s.

 \begin{figure}[!htp]
 \center
 \psfig{figure=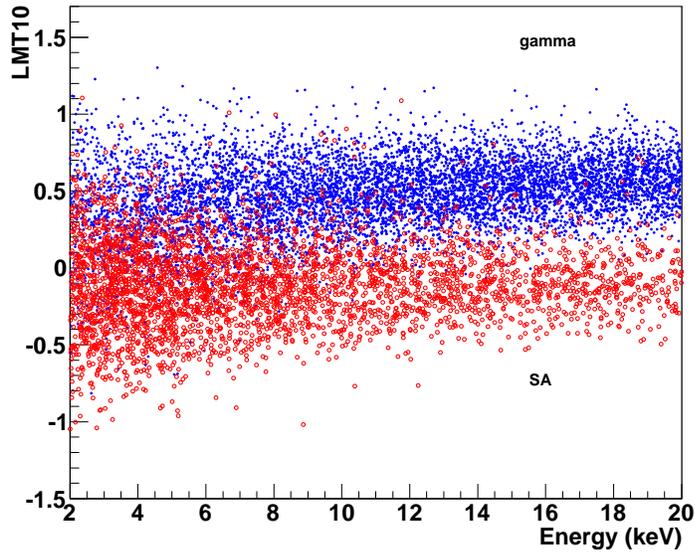, angle =0,width = 4.0 in}
\caption{LMT10 of SA events and gamma events in RDD} \label{logrmt10sa}
\end{figure}

\section{Mean-time distribution of SA events}    
   
  In the KIMS experiment, the quantity LMT10 is used as the PSD discriminator.
  The distribution of LMT10 for SA events must be understood in order to distinguish them from dark matter
  candidate events in KIMS. For this we analyzed the response of crystal A in the SA-tagged events 
  selected according to the description in the previous section.

 \begin{figure}[!htp]
\center \psfig{figure=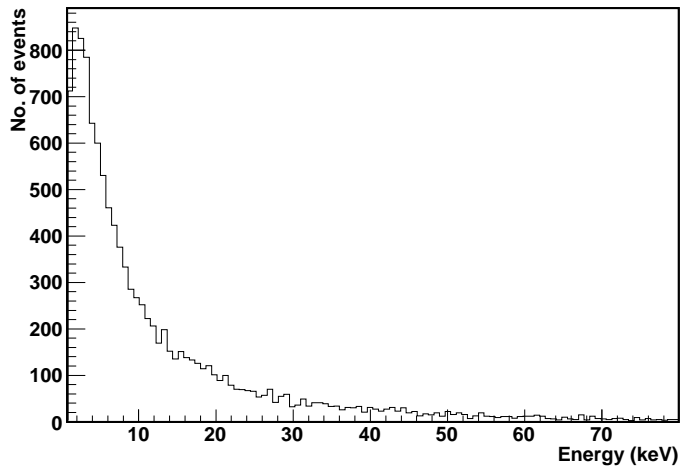, width = 4
in}
\caption{The energy spectrum in crystal A for SA-tagged events. 
 } \label{Esa}
\end{figure}

  Figure~\ref{logrmt10sa} shows the LMT10 distribution for SA-
  and gamma ray-induced events from RDD in the low energy range that is used in the dark matter search. 
  The SA events range down to 3 keV, the energy threshold for this study. Figure~\ref{Esa} shows the energy spectrum of SA events. 
  The energy deposited in crystal A is the sum of the recoil energy of the recoiling \Pbc~ nucleus and some partial energy of the alpha particle.      
  The \Pbc~ recoil energy is 103 keV, and it is expected to show up as 7--8 keV due to the quenching factor of \csitl if we assume the
  quenching factor for \Pbc~ is the same as that for Cs and I nuclei~\cite{qfactor1,qfactor2}. 
  The FWHM energy resolution at this energy is 4 keV.
  There is no clear \Pbc~ signal such as that which was seen in another study that used a phonon sensor~\cite{cresstA}. 
  The scintillator may have an inactive surface layer where the scintillation efficiency is very small that
  it can cause smearing of \Pbc~ signal.

\begin{figure}[!htp]
\center \psfig{figure=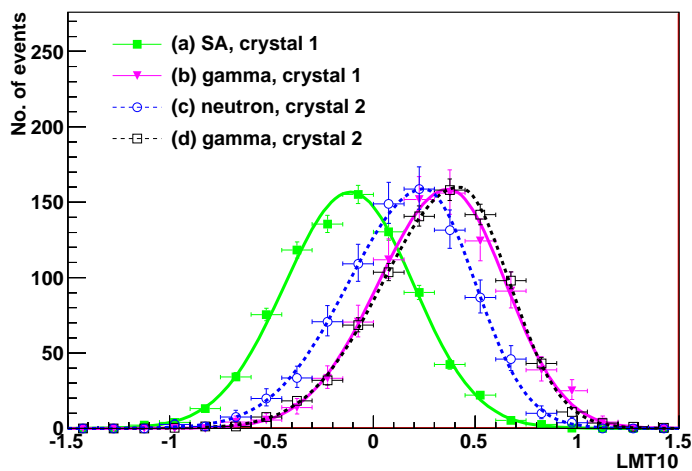, width = 4
in}
\caption{ LMT10 distibutions for various reference data at 3 keV. (a) LMT10 for SA-induced events in crystal 1, (b) gamma-induced events in crystal 1, (c) neutron-induced events in crystal 2 and (d) gamma-induced events in crystal 2.} 
 \label{complogrmt10b}
\end{figure}

\begin{figure}[!htp]
\center \psfig{figure=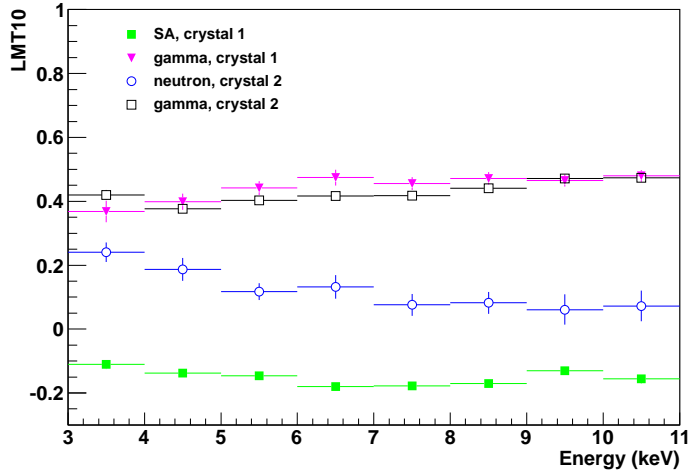, width = 4
in}
\caption{
The value of the peak position of LMT10 distributions for SA-, neutron- and  gamma-induced events. 
Crystal 1 refers to the crystal used for SA study and crystal 2 
is the one used for the neutron response measurement.
} \label{complogrmt10a}
\end{figure}

  The LMT10 distributions for 3--4 keV energy bin
  are shown in Fig.~\ref{complogrmt10b}.  
  The crystal used for the SA study with \rn~ contamination is labeled crystal $1$ in the figure; 
  the crystal used for the neutron response study is a different crystal, crystal 2, which was exposed to an Am-Be neutron source.
  Here, we also present the results of gamma calibration for both test crystals for comparison. 
  The gamma calibration was done by irradiating the crystals with \cs~ source.
  The temperature for SA study setup was maintained as $(25.4\pm 0.3)^{\circ}\mathrm{C}$, and for neutron study setup, $(25.3\pm 0.7)^{\circ}\mathrm{C}$.
  The results in the
  figure show that SA-induced signals are, on average, faster than neutron-induced signal events, and their LMT10
  distributions are distinct.   
  Figure~\ref{complogrmt10a} shows the value of the peak position of LMT10 distribution as a function of the energy for each type of event, which is not
  necessarily same with the mean value since the distribution is an asymmetric Gaussian. 
  The SA LMT10 distribution obtained from this study is used to distinguish possible WIMP-induced nuclear recoils
  from surface alpha-induced events in the KIMS data.

\section{Conclusion}

 We have studied surface alpha background from \rn~ progenies with a \csitl scintillator contaminated
 with  \rn~ progenies, mainly \Pbb~ and its daughters. The main alpha emitter is \Poc.
 When \Poc~ alpha decays at the surface, it doesn't deposit its full energy into the detector. Its energy spectrum 
 ranges from the full peak energy to the very low energy. We directly show that SA events
 decay faster than neutron and gamma events. We also obtained the distribution of LMT10, the PSD parameter, for SA events. This can be used to analyse the dark matter search data of KIMS experiment.   

\section*{Acknowledgments}
 We thank Dr. Jongman Lee for providing us access to the KRISS radon chamber.
 We are very grateful to the Korea Midland Power Co. and their staff for
 providing the underground laboratory space at Yangyang.
 This research was supported by the WCU program (R32-10155) and Basic Science Research Program (2010-0005332)  through National Research Foundation of Korea funded by the Ministry of Education, Science and Technology.

\end{document}